\documentclass[aps, twocolumn, showpacs, superscriptaddress,floatfix]{revtex4-1}
\usepackage{graphicx,amsmath,amssymb}
\usepackage{bm}
\usepackage{braket}
\usepackage{dsfont}
\begin{document}
\title{Exact Topological Flat Bands from Continuum Landau Levels}
\author{Junkai Dong}
\email{jd722@cornell.edu}
\author{Erich J. Mueller}
\email{em256@cornell.edu}
\affiliation{Laboratory of Atomic and Solid State Physics, Cornell University, Ithaca NY 14853}

\begin{abstract}
We construct and characterize
tight binding Hamiltonians which contain a completely flat topological band made of 
continuum lowest Landau level wavefunctions sampled on a lattice. 
We find an infinite family of such Hamiltonians, with simple analytic descriptions.  These provide a valuable tool for constructing exactly solvable models.  We also implement a numerical algorithm for finding the most local Hamiltonian with a flat Landau level.  We find intriguing structures in the spatial dependence of the matrix elements for this optimized model.  
The models we construct serve as foundations for numerical and experimental studies of topological systems,
both non-interacting and interacting.
\end{abstract}
\maketitle
\section{Introduction}
Over the past decade there has been intense interest in model Hamiltonians which contain
flat bands with non-trivial topological indices \cite{reviewstop}.  When fermions partially fill a flat band, interactions become the most important scale, and inevitably lead to non-trivial strongly correlated states of matter. In topologically trivial flat bands, these are typically charge density waves or Wigner crystals.
In topologically non-trivial flat bands, they are even more exotic.
The best example is the fractional quantum Hall effect \cite{fqh}.  Here we construct a family of particularly interesting topologically non-trivial flat band models, and explore their properties.

The prototypical topological flat band model is simply 2D electrons in a uniform magnetic field.  The bands there are Landau levels.  One elegant feature of this system is that one can choose a gauge where the lowest energy Landau level (LLL) is spanned by wavefunctions which have the form of an arbitrary analytic function times a Gaussian envelope.
Inspired by the work of Kapit and Mueller \cite{km}, which built upon results in \cite{Perelomov1971,PhysRevLett.99.097202}, we consider lattice models where the flat band is spanned by exactly the same wavefunctions. As noted by Ataki\c{s}i and Oktel, the space of such Hamiltonian can be constructed via a projector technique \cite{oktel}.  We find two important results regarding these flat band models: (1) We analytically construct a family of Hamiltonians in this space, generalizing the singular example in \cite{km}, and extending it to more general lattices.  (2) We numerically find the most local Hamiltonian in this space. 

In addition to being fundamental to our understanding of topological band structures, this construction is of practical value.  For example, Kapit, Ginsparg, and Mueller \cite{kgm} used the results from \cite{km} to construct a numerical technique for studying the braiding of fractionally charged excitations with non-trivial statistics.  The new examples presented here can be used for similar purposes.  Additionally, these Hamiltonians can inspire the design of artificial structures such as lattices of coupled Josephson junctions, or optical lattices.  Of particular value, the present work finds the most localized Hamiltonian with a flat lowest Landau level, an important consideration given
the challenges of implementing long range couplings.
Furthermore, our flat band models can provide settings in which the many-body problem can be solved exactly \cite{km}. These kinds of flat bands give rise to fractional Chern insulators, which are studied both theoretically \cite{PhysRevX.1.021014,PhysRevLett.106.236804,PhysRevLett.108.126805,PhysRevLett.110.185302,PhysRevLett.112.156801} and experimentally \cite{Spanton62}. Our Hamiltonians can also be translated into spin liquid language and give rise to interesting physics similar to that in \cite{PhysRevLett.99.097202,PhysRevB.80.104406,PhysRevB.89.165125}.

In section~\ref{construct} we give our analytic construction of a two-parameter family of Hamiltonians which contain a flat topological band spanned by continuum lowest Landau level wavefunctions.  Section~\ref{gen} explores the more general problem of finding the most local Hamiltonian with this property.  Conclusions and outlook are in section~\ref{conc}.

\section{Construction for flat band hamiltonians}\label{construct}
Here we construct a family of topological tight-binding models, which describe the motion of a single electron
hopping between sites in the 2D plane.
The most general Hamiltonian of this form can be written
\begin{equation}
H=\sum_{i,j}J(z_i,z_j)|i\rangle\langle j|,
\end{equation}
where $|j\rangle$ is the state where the electron is at site $j$, located at position $z_j \ell =(n_j w_1+m_jw_2)\ell$, where $\ell$ is the length unit which can be chosen arbitrarily.  The dimensionless complex numbers $w_1$ and $w_2$ represent the generators of the lattice, and $n_j,m_j$ are integers.  We require that the generators are not linearly dependent, ie. ${\rm Im}(w_1^* w_2)\neq 0$.  In section~\ref{moregeometries} we give some
discussion of the non-Bravais case where there is more than one site per unit cell. By definition $J(z_i,z_j)=\braket{i|H|j}$.

We wish to choose the hopping matrix elements $J(z_i,z_j)$ such that the 
nullspace of $H$ contains all states of the form
 $|\psi_\nu\rangle=\sum_{j}\psi_{\nu j} |j\rangle$ with
\begin{equation}\label{res1}
\psi_{\nu j}=\braket{j|\psi_\nu}=z_j^\nu\exp(- \frac{\pi \phi}{2}|z_j|^2).
\end{equation}
These define the lowest Landau level of the continuum problem.  The length-scale in these wavefunctions is set by the parameter $\phi$, which corresponds to the strength of the magnetic field -- in particular, $\phi=eB \ell^2/h$ is the magnetic flux through one unit area.  This construction will yield a flat band with eigen-energy $0$.  We will describe a Hamiltonian with this flat-band property as a ``parent Hamiltonian'' for the lowest Landau level.

By construction, this band will have Chern number 1 and if filled will display a quantized Hall effect. The argument is simply the one given by
Thouless, Kohmoto, Nightingale and den Nijs \cite{tknn} for the continuum.  Their argument only relies upon the wavefunctions, and not the underlying Hamiltonian.

In \cite{km}, Kapit and Mueller used the ``singlet sum rule" \cite{ssr}, generalized in \cite{PhysRevB.85.155145}, to find one such set of hopping matrix elements in the case of a square lattice: $w_1=1, w_2=i$.
In particular, for any analytic function $f(z)$, the following sum vanishes
\begin{equation}\label{ssr}
\sum_{z=n+im
}(-1)^{m+n+mn+1}f(z)\exp(- \frac{\pi}{2} |z|^2)=0,
\end{equation}
where $n$ and $m$ are integers.
This identity emerges from the Poisson sum,
\begin{equation}\label{expsum}\sum_{z=n+im}\exp(cz- \frac{\pi}{2} |z|^2)=2\sum_{z=2n+2im}\exp(cz- \frac{\pi}{2} |z|^2).
\end{equation}
Subtracting the two sides of the equation, differentiating with respect to $c$ an arbitrary number of times, then setting $c=0$ yields Eq.~(\ref{ssr}).
Consequently, if one takes
\begin{equation}\label{origin}
J(z_i,z_j)=(-1)^{n+m+nm+1}e^{-\frac{\pi(1- \phi)}{2}|z|^2}e^{i\pi \phi{\rm Im}(z_iz^*)}\end{equation}
in which $z=z_j-z_i=n+im$, then the functions in Eq.~(\ref{res1})
will be in the nullspace of $H$.  Note that the prefactor $(-1)^{n+m+nm}$ has an alternating pattern with period 2 in each direction.  We will produce models with other periods of modulation, while extending it to
 generic Bravais lattices.
\par
Note that Eq.~(\ref{expsum}) can be interpreted as an identity of Jacobi theta functions:
\begin{equation}
\begin{aligned}
&\sum_{z=n+im}\exp(cz-\frac{\pi}{2}|z|^2)=\theta_3(\frac{c}{2},e^{-\pi/2})\theta_3(\frac{ic}{2},e^{-\pi/2})\\=&2\theta_3(c,e^{-2 \pi})\theta_3(ic,e^{-2\pi})=2\sum_{z=2n+2im}\exp(cz- \frac{\pi}{2} |z|^2).
\end{aligned}
\end{equation}
in which the Jacobi theta function $\theta_3$ is defined as
\begin{equation}
\theta_3(z,q)=\sum_{n\in \mathbb{Z}}q^{n^2}\exp(2inz)
\end{equation}
In the above construction, the key is Eq.~(\ref{expsum}).  Consider $\mathcal{L}=\{n_j w_1+m_jw_2\}$, where as before $w_1$ and $w_2$ represent the generators, and $n_j,m_j$ are integers.  We define the sublattice $\mathcal{L}_k=\{k n_j w_1+k m_jw_2\}$.  Below we prove that for all complex numbers $c$,
%
\begin{eqnarray}\label{id1}
\sum_{z\in \mathcal{L}}\exp\left(-\frac{\pi}{k\Omega}|z|^2\right)\exp(cz)=\qquad \\\nonumber
\qquad k\sum_{z\in \mathcal{L}_k}\exp\left(-\frac{\pi}{k\Omega}|z|^2\right)\exp(cz)
\end{eqnarray}
where $\Omega={\rm Im}(w_1^* w_2)$ is the area of the unit cell.
Using this identity, following our previous argument,
 the wavefunctions in  Eq.~(\ref{res1})  will be in the null-space of $H$ as long as
\begin{equation}\label{sol}
\begin{aligned}
J_k(z_i,z_j)=f_k(z) \exp(\frac{\pi\phi}{2}|z|^2+i \pi \phi{\rm Im}(z_iz^*))\\
f_k(z)=G_k(z)\exp(-\frac{\pi}{k\Omega}|z|^2)
\end{aligned}
\end{equation}

in which $z=z_j-z_i=nw_1+mw_2$, and $G_k(z)$ is $1$ unless $z\in{\mathcal{L}_k}$, in which case $G_k(z)=1-k$.
For the special case of a square lattice with $\Omega=1$ and $k=2$ we reproduce Eq.~(\ref{ssr}).
\par


To prove Eq.~(\ref{id1}),
%
we again use the Poisson summation formula, which
 for an arbitrary Bravais lattice $\mathcal L$ in 2D  reads \begin{equation}\label{poisson}
\sum_{z\in \mathcal{L}}\exp(cz- \frac{1}{2}|z|^2)=\frac{2 \pi}{\Omega}\sum_{z\in \mathcal{L}^*}\exp(icz- \frac{1}{2} |z|^2).
\end{equation}
Here 
$\mathcal{L}^*$ is the dual lattice, generated by the reciprocal lattice vectors $u_1,u_2$ satisfying
$\mathop{Re}(w_i^* u_j)=2\pi \delta_{ij}$.
\par
To simplify the subsequent notation, we rescale $w_1$ and $w_2$ so that the unit cell of the reciprocal lattice has an area which is an integer multiple of $2\pi$. That is, the size of unit cell is $\Omega=2 \pi/k$ in which $k\in \mathbb{Z}$. Hence
$u_1=-i k w_2$ and $u_2=i k w_1$.
Note that the reciprocal lattice $\mathcal{L}^*$ is simply a rotation of $\mathcal{L}_k$:
if $u\in \mathcal{L}^*$ is an element of the reciprocal lattice, then $i u$ is in $\mathcal{L}_k$.
%
The Poisson summation formula can
then be written:
%
\begin{equation}\label{poi2}
\sum_{z\in {\cal L}}\exp(cz-\frac{1}{2}|z|^2)
=k\sum_{z\in {\cal L}_k}\exp(cz-\frac{1}{2}|z|^2).
\end{equation}
Rescaling back to the original lattice vectors yields Eq.~(\ref{id1}).

\subsection{Non-Bravais lattices}\label{moregeometries}
The results from the previous section can also be extended to the non-Bravais case.

Consider a lattice with a $n$-point basis.  It is defined by
an underlying Bravais lattice $\mathcal{L}$ generated by $w_1$ and $w_2$, and basis
vectors described by the
complex numbers ${v}_0=0,{v}_1,\dots, {v}_{n-1}$, so that the points in the lattice are of the form
$z=a w_1+b w_2+v_j$, where $a$ and $b$ are integers.
Motivated by our previous discussion, we rewrite the hopping matrix element to express the Hamiltonian as
\begin{equation}
\begin{aligned}
H=\sum_{i,j}\sum_{z,w\in \mathcal{L}}& f_{ij}(z_{ij})\exp(i \pi \phi{\rm Im}((w+{v}_j)z_{ij}^*))\\&\exp(\frac{\pi \phi}{2}|z_{ij}|^2)\ket{z+{v}_i}\bra{w+{v}_j}
\end{aligned}
\end{equation}
in which $z_{ij}=w+{v}_j-z-{v}_i$.
As should be apparent from the notation, the indices $i,j$ run from $0$ to $n-1$, and they index the basis vectors.
  We wish to choose $f_{ij}$ such that $H\ket{\psi_\nu}=0$ for all $\nu$ -- a
condition which is equivalent to requiring that the following $n$ constraints are satisfied for all complex numbers $c$:
\begin{equation}\label{nonb}
\begin{aligned}
&\sum_{z\in \mathcal{L}}\sum_{i}f_{0i}(z+{v}_i-{v}_0)\exp(c(z+{v}_i-{v}_0))=0\\
&\dots\\
&\sum_{z\in \mathcal{L}}\sum_{i}f_{(n-1)i}(z+{v}_i-{v}_{n-1})\exp(c(z+{v}_i-{v}_{n-1}))=0.
\end{aligned}
\end{equation}
A trivial way to
satisfy these constraints is to separately satisfy the following $n^2$ constraints:
\begin{equation}\label{individual}
\sum_{z\in \mathcal{L}}f_{ji}(z+{v}_i-{v}_j)\exp(c(z+{v}_i-{v}_j))=0.
\end{equation}
Each of these are equivalent to our original Bravais lattice problem, and hence we can take $f_{ij}$ to be given by Eq.~(\ref{sol}),
\begin{equation}
f_{ji}(z+{v}_i-{v}_j)=f_k(z)
\end{equation}
for arbitrary integer $k$.\par
Additionally, we can construct solutions of Eq.~(\ref{nonb}) which do not also satisfy Eq.~(\ref{individual}).  For example, we can take
\begin{equation}\label{basis1}
f_{ij}(z)=\exp(-\frac{\pi}{k \Omega}|z|^2)
\end{equation}
for $i\not=j$,
and
\begin{equation}\label{basis2}
f_{ii}(z)=G_{ik}(z)\exp(-\frac{\pi}{k \Omega}|z|^2)
\end{equation}
in which $z=z_j-z_i=nw_1+mw_2$. $G_{ik}(z)$ is slighly modified from $G_k(z)$ defined before: $G_{ik}(z)$ is $1$ unless $z\in{\mathcal{L}_k}$, in which case $G_{ik}(z)=1-k\sum_i\exp(i\frac{2 \pi}{k \Omega}{\rm Re}(z^*v_i))$.
To show that this expression sattisfies Eq.~(\ref{nonb}), we
%
shift
the lattice vectors in Eq.~(\ref{poisson}) to get
%
\begin{equation}
\sum_{z\in \mathcal{L}}f(z+{v})=\frac{2 \pi}{\Omega}\sum_{z\in \mathcal{L}^*} \tilde{f}(z)\exp(i {\rm Re}(z^* {v}))
\end{equation}
in which $\tilde{f}$ is the fourier transform of $f$.
Summing over the basis vectors then yields the identity
\begin{equation}
\sum_{z\in \mathcal{L}}\sum_if(z+{v}_i)=\frac{2 \pi}{\Omega}\sum_{z\in \mathcal{L}^*}\tilde{f}(z)(\sum_i\exp{(i {\rm Re}(z^* {v}_i))})
\end{equation}
\par
As before, we rescale the lattice so that $\mathcal{L}^*$ is a subset of $\mathcal{L}$, and take
take $f(z)=\exp(cz-\frac{1}{2}|z|^2)$, $\tilde{f}(z)=\exp(icz-\frac{1}{2}|z|^2)$.  We then subtract the two sides, to get Eq.~(\ref{basis1}), (\ref{basis2}).\par

\begin{figure}[t]
\includegraphics[scale=0.5]{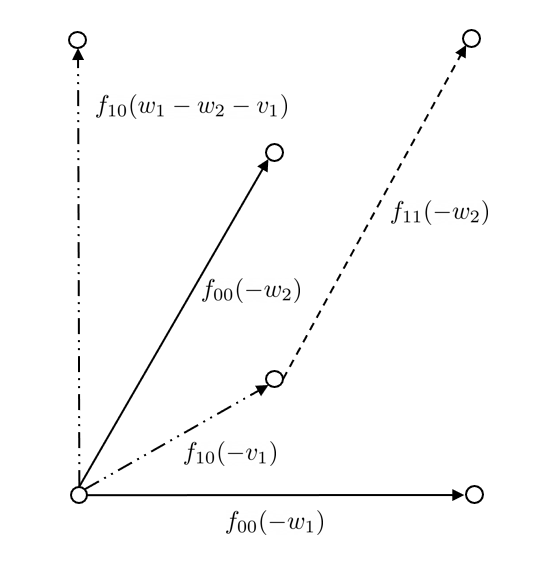}
\caption{Graphical depiction of the matrix elements on a honeycomb lattice.
Solid arrows: hoppings from sublattice $0$ to sublattice $0$. Dashed arrows: hoppings from sublattice $1$ to sublattice $1$. Dash-dotted arrows: hoppings between sublattices. $w_1,w_2$ are the lattice generators while $v_1$ is the vector connecting the two sites in a unit cell.}
\label{fig:hophoneycomb}
\end{figure}

One concrete application of this procedure is constructing a flat topological band on a honeycomb lattice.
In particular, we take $w_1=\sqrt{2}/\sqrt[4]{3},w_2=1/\sqrt[4]{12}(1+i\sqrt{3}))$, ${v}_1=1/\sqrt[4]{12}(1+i/\sqrt{3})$. The simplest matrix elements are: \begin{equation}
\begin{aligned}
&f_{00}(z)=-\exp(-\pi|z|^2)\exp(2 \pi i {\rm Re}(z^* {v}))\\
&f_{11}(z)=-\exp(-\pi|z|^2)\exp(-2 \pi i {\rm Re}(z^* {v}))\\
&f_{01}(z)=f_{10}(z)=\exp(-\pi|z|^2).
\end{aligned}
\end{equation}
If we truncate this Hamiltonian to only nearest and next-nearest neighbor hopping at $\phi=1$, it reduces to the Haldane model \cite{haldane}.
The hoppings in the $\phi=1$ tight binding Hamiltonian is shown in Fig. \ref{fig:hophoneycomb}.

\section{Most Localized Parent Hamiltonian}\label{gen}

Here we numerically construct the most localized Hamiltonian with the property that the continuum lowest Landau wavefunctions form a flat band with $E=0$. The construction is meaningful since it has been shown that the Hamiltonians with the property cannot be local; they must contain arbitrarily long range hoppings \cite{Chen_2014}.  Nonetheless, as the Kapit-Mueller Hamiltonian demonstrates, the matrix elements can fall off at least as fast as a Gaussian. Note that we are not restricting to Hamiltonian of the form of section II. Rather we are looking at completely general Hamiltonians that annihilate the lowest Landau level.
\subsection{Formalism}
Let $P$ be the projector into the lowest Landau level, and $\bar{P}=\mathds{1}-P$ be the projector into the orthogonal space.
A parent Hamiltonian has the property $\bar{P}H\bar{P}=H$.
We define the range of $H$ via
%
\begin{equation}
R^2=\frac{\sum_{i,j}|\braket{i|H|j}|^2|z_i-z_j|^2}{\sum_{i,j}|\braket{i|H|j}|^2}=\frac{\sum_{i,j}|J(z_i,z_j)|^2|z_i-z_j|^2}{\sum_{i,j}|J(z_i,z_j)|^2}
\end{equation}
We want to minimize this expression over all parent Hamiltonians which are invariant under magnetic translations:
i.e. $\braket{i|H|j}\exp(- i\phi_{AB}(z_i,z_j))=$ $\braket{i-j|H|0}$ $\exp(-i\phi_{AB}(z_i-z_j,0))$, in which $\phi_{AB}(z_i,z_j)$ $=\pi \phi {\rm Im}(z_i z^*)$ is the Aharanov-Bohm phase associated with direct motion from $j$ to $i$. Here $z=z_j-z_i$. Due to this symmetry
the range can be expressed
as
\begin{equation}
\begin{aligned}
R^2&=\frac{\sum_{j}|\braket{j|H|0}|^2|z_j|^2}{\sum_{j}|\braket{j|H|0}|^2}=\frac{\sum_{j}|J(z_j,0)|^2|z_j|^2}{\sum_{j}|J(z_j,0)|^2}\\
&=\frac{\bra{0}H r^2 H\ket{0}}{\bra{0}H^2\ket{0}},
\end{aligned}
\end{equation}
where $r^2= \sum_j\ket{j}|z_j|^2\bra{j}$.  This expression can be further simplified by defining the wavefunction
$\ket{\psi_H}=H\ket{0}$, in terms of which $R^2= \braket{\psi_H|r^2|\psi_H}/\braket{\psi_H|\psi_H}$.  We wish to minimize $R^2$ with
respect to $\ket{\psi_H}$ with the constraint that $\ket{\psi_H}$ is in the image of $\bar{P}$, i.e. $\ket{\psi_H}$ is orthogonal to the space spanned by lowest Landau level wavefunctions. For the resulting Hamiltonian to be Hermitian, we also require $\langle i | \psi_H\rangle=\langle -i|\psi_H\rangle^*$ for all $z_i$ in the lattice.  We denote the projector into the space obeying this latter constraint as $P^\prime$ and the projector into the space obeying both constraints as $\mathbb{P}$.  Our minimization problem is then
equivalent to finding the smallest non-zero eigenvalue of $\mathbb{P}r^2\mathbb{P}$.  The eigenvector's components, $\braket{j|\psi_H}=\braket{j|H|0}$, correspond to matrix elements of the Hamiltonian: the remaining matrix elements can be found by using magnetic translations.

We work on a finite $L\times L$ square lattice with periodic boundary conditions.  These boundary conditions are only well behaved if the total flux through the lattice is an integer.  Choosing the lattice spacing to be unity, this corresponds to requiring $\phi L^2$ to be an integer.  We define $\phi=p/q$.

We explored a number of ways of constructing the projector $\bar P$, and found that when the denominator $q$ is small,
the most numerically efficient approach involved producing  the Kapit-Mueller Hamiltonian $H_{KM}$.  We numerically found its eigenstates, then used them to produce the projector into the lowest Landau level, and its complement $\bar P$ as  $L^2\times L^2$ matrices. In this construction we use periodic boundary condition. However, as we explain below, we systematically study different system sizes and find that our results are independent of $L$ for sufficiently large $L$.

One technical issue is that $P^\prime$, which projects into the space where $\braket{i|\psi_H}=\braket{-i|\psi_H}^*$, can only be represented as a linear operator if we work in an enlarged space, considering the length $2L^2$ vector with components $\langle 1 |\psi_H\rangle,\ldots \langle L+iL|\psi_H\rangle, \langle 1|\psi_H\rangle^*,\ldots\langle L+iL|\psi_H\rangle^*$. $P^\prime$ is then the $2L^2\times 2L^2$ matrix made of four $L^2\times L^2$ blocks
\begin{equation}
P^\prime=\frac{1}{\sqrt{2}}\left(\begin{array}{cc}I&Q\\Q^T&I\end{array}\right)
\end{equation}
Here $I$ is the identity matrix while $Q$ is a permutation matrix: its non-zero entries connect the elements $\braket{i|\psi_H}$ and $\braket{-i|\psi_H}^*$.

In this larger space, $\bar P$ is just a block matrix, where the two blocks are the previously constructed $\bar P$ and its complex conjugate.  The mutual projector is constructed as $\mathbb{P}=2 P^\prime (P^\prime+\bar P)^{-1}\bar P$, where $(\cdots)^{-1}$ denotes the pseudoinverse \cite{proj}.  We use standard packages to numerically calculate the pseudoinverse.
Matrix multiplication then gives $\mathbb{P}r^2 \mathbb{P}$.  Numerically diagonalizing this matrix is straightforward.

%
\subsection{Results}
\begin{figure}[t]
\includegraphics[scale=0.5]{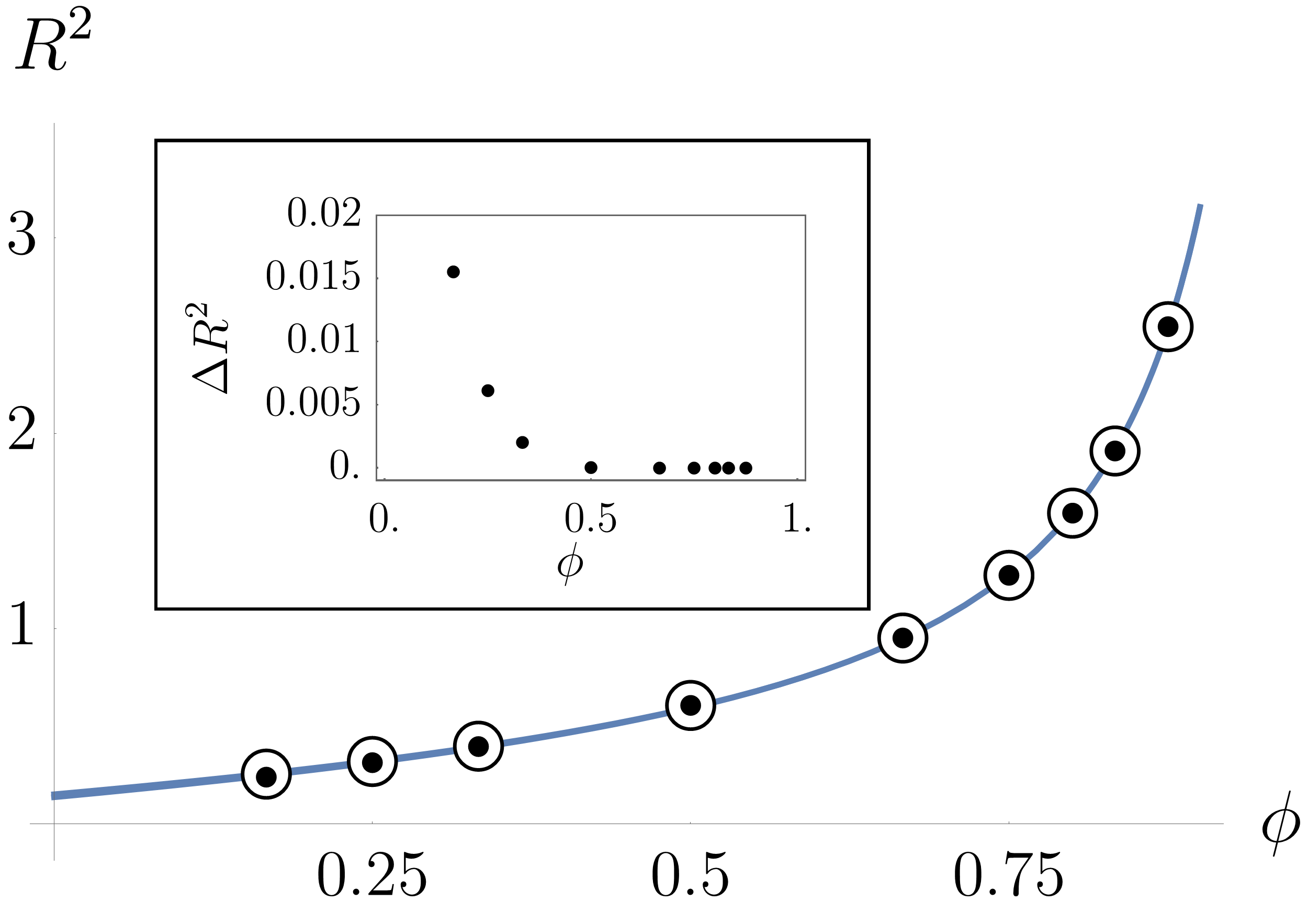}
\caption{Range $R^2$ of the parent Hamiltonian versus magnetic field strength $\phi$.
Open Circles: $R^2_{KM}$, corresponding to Eq.~(\ref{origin}). Solid Dots: $R^2$ of the most localized parent Hamiltonian.
Line is an empirical fit to the solid dots, Eq.~(\ref{ft}).  Inset: deviation $\Delta R^2= R^2_{KM} - R^2$.  }
\label{rsq}
\end{figure}
Figure~\ref{rsq} shows the range of the Hamiltonian, $R^2$, as a function of the magnetic field strength.  
In order to have a commensurate flux, different $L$ are used for different $\phi$: the data shown corresponds to 
$(\phi,L)=(\frac{1}{6},72)$, $(\frac{1}{4},48)$, $(\frac{1}{3},36)$, $(\frac{1}{2},24)$, $(\frac{2}{3},36)$, $(\frac{3}{4},48)$, 
$(\frac{4}{5},40)$, 
$(\frac{5}{6},42)$, $(\frac{7}{8},48)$. At each of these $\phi$, we varied $L$, and verified that finite size effects were negligible.  The main feature of the data is that
the range monotonically increases with $\phi$,  diverging as $\phi\to 1$.  As shown in the figure, the curve
is well approximated by
\begin{equation}\label{ft}
R^2=-a+\frac{b}{1- \phi}+c\phi
\end{equation}
with $a\approx -0.168263,b\approx 0.305039,c\approx 0.303745$.

\begin{figure}[tp]

\includegraphics[width=0.9\linewidth]{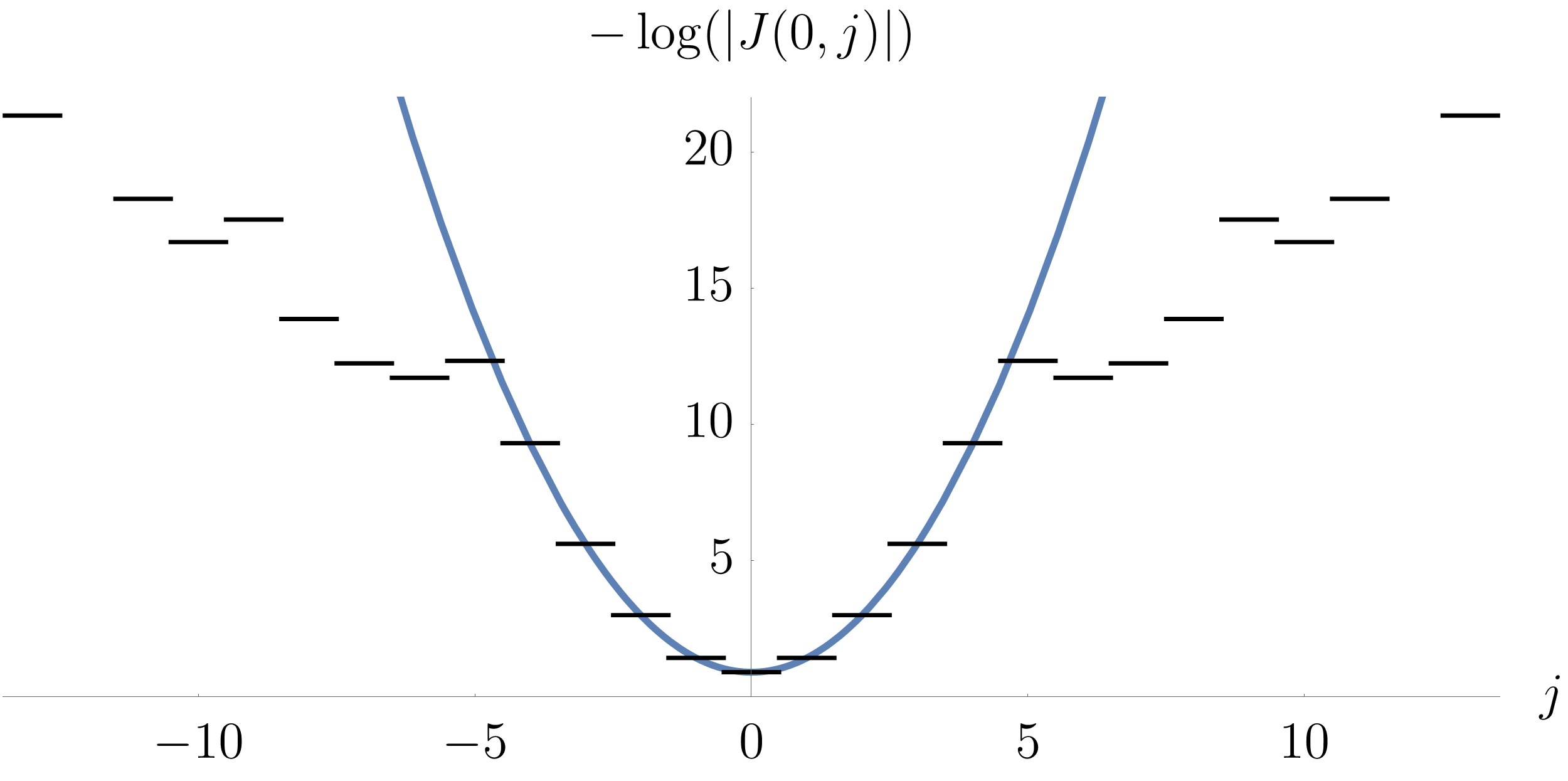}\\(a) $\phi=2/3$\\

\includegraphics[width=0.9\linewidth]{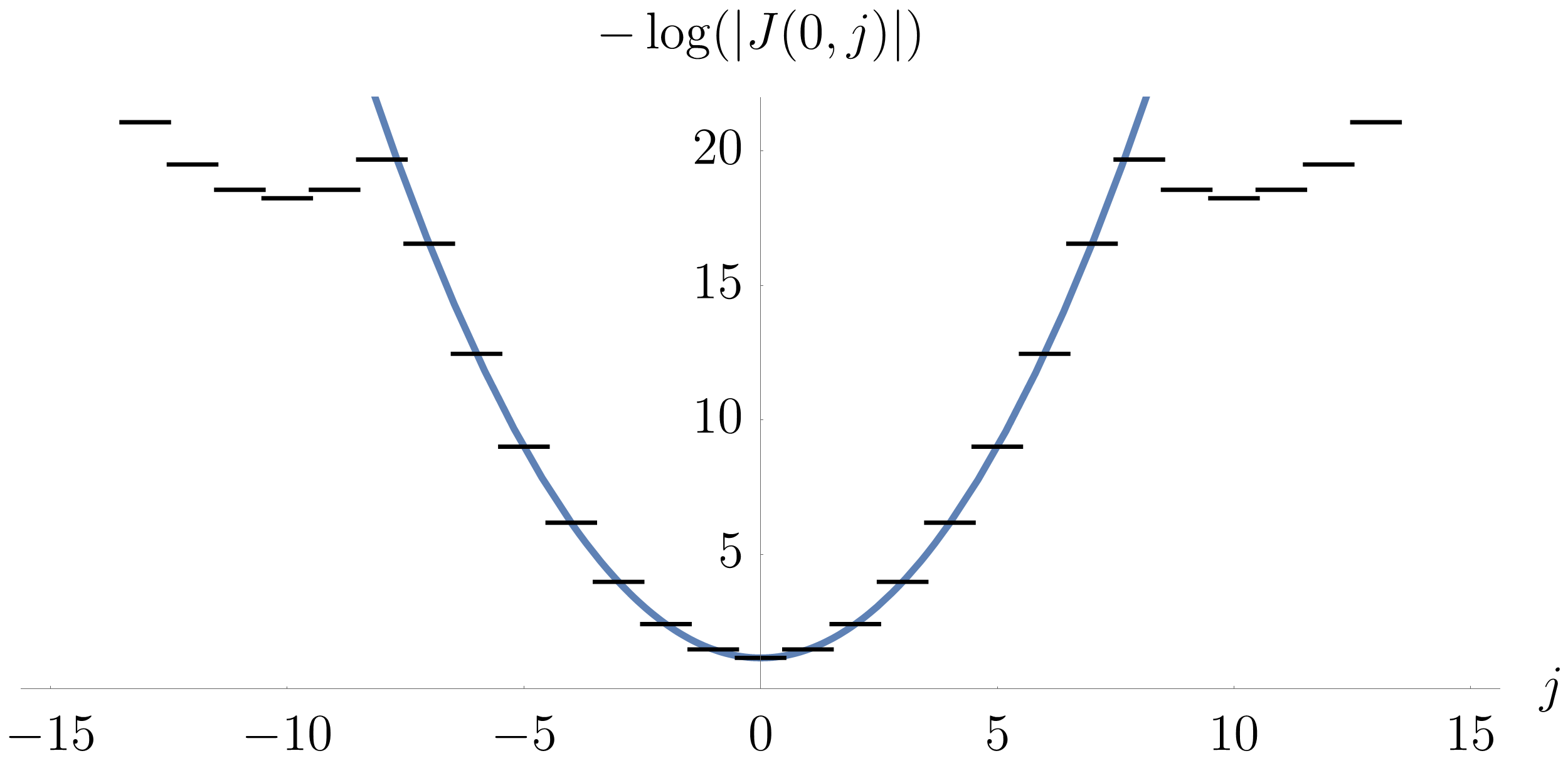}\\(b) $\phi=4/5$\\

\includegraphics[width=0.9\linewidth]{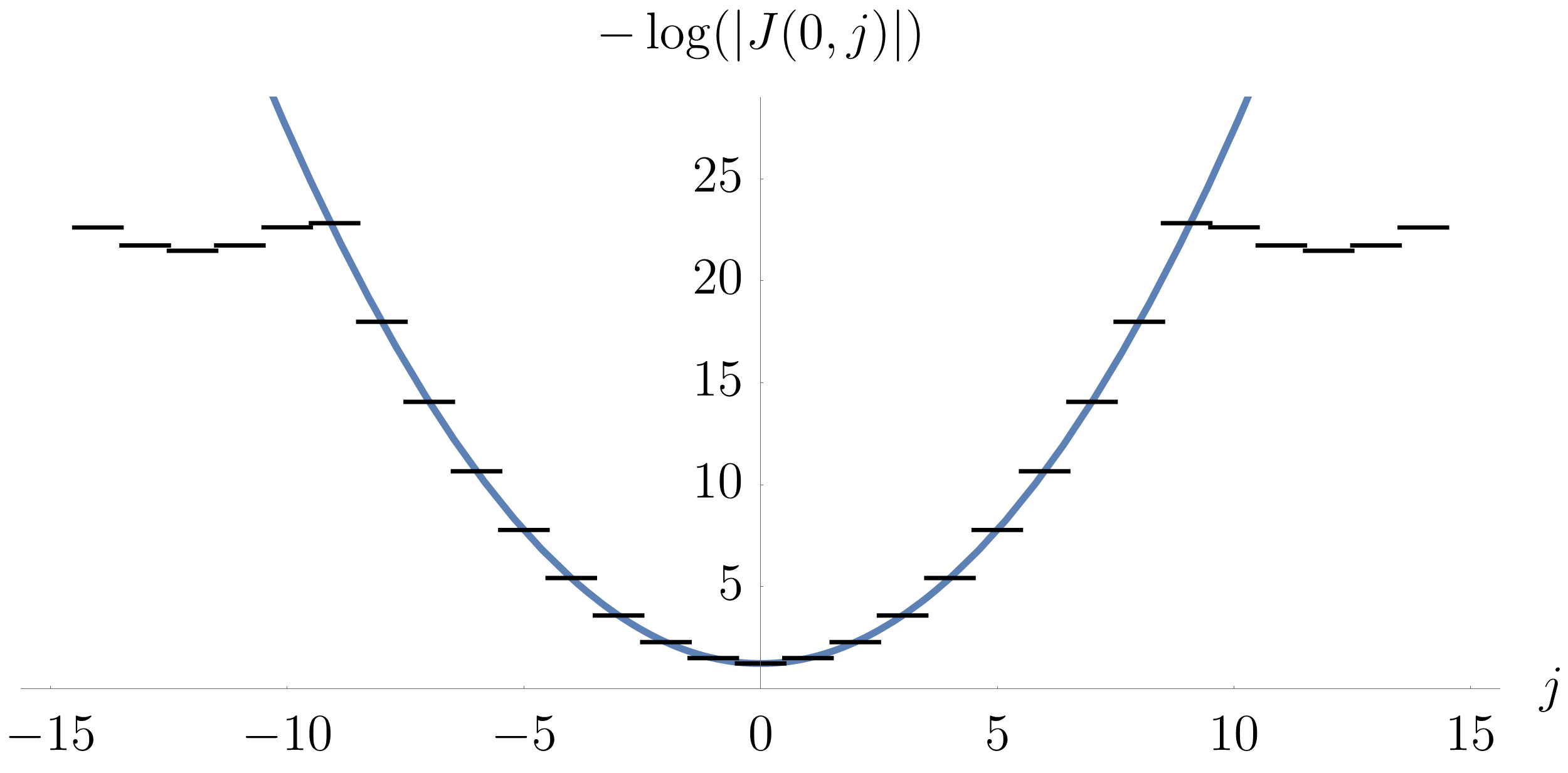}\\(c) $\phi=5/6$\\
\caption{Hopping magnitude $-\log|J(0,j)|$ for the most local tight binding model with a flat Lowest Landau Level.  The magnetic field corresponds to (a) $\phi=2/3$, (b) $\phi=4/5$, (c) $\phi=5/6$.  The solid line shows the analytic expression for $-\log|J_{KM}|$.}
\label{fig:hops}
\end{figure}

We find that the Kapit-Mueller Hamiltonian in Eq.~(\ref{origin}) nearly saturates our numerical bound: the open circles in Fig.~\ref{rsq} show the Kapit-Mueller result, and the inset shows the deviation between the two.  Not only is the difference
 consistently small, but when $\phi$ exceeds $0.5$, the deviation exceeds machine precision. Clearly the Kapit-Mueller Hamiltonian is a good aproximation of the most localized Hamiltonian. We emphasize, however, that for $\phi<1/2$ the Kapit-Mueller Hamiltonain clearly has a longer range than the optimal Hamiltonian.\par
\par
Given the close agreement, we can gain some analytic understanding of Eq.~(\ref{ft}) by analyzing the range of the Kapit-Mueller Hamiltonian,
$R^2_{KM}$, which can be expressed as \begin{equation}
R^2_{KM}=\frac{\sum_{x,y}(x^2+y^2)\exp(-(1- \phi)\pi(x^2+y^2))}{\sum_{x,y}\exp(-(1- \phi)\pi(x^2+y^2))}.
\end{equation}

In the $\phi\to 1$ limit the sum can be replaced by an integral, yielding
%
\begin{equation}
\lim_{\phi\to 1^{-}}R_{KM}^2
=\frac{1}{\pi(1- \phi)},
\end{equation}
where $1/ \pi\approx0.31831$ is very close to the coefficient $b$ in our fit.\par

Given this agreement, it is not surprising that the matrix elements of the optimized Hamiltonian are related to those of Eq.~(\ref{origin}).  The similarity is particularly striking at short distances.  

Figure~\ref{fig:hops} shows the logarithm of the magnitude of the hopping matrix elements in the horizontal direction for different values of $\phi$.  As is apparent, $-\log|J(0,j)|$ is made up of a sequence of parabolas, implying that $|J(0,j)|$
is well described by a discontinuous set of Gaussians.  Comparison with the solid line indicates that the lengthscale of the central Gaussian is the same as Eq.~(\ref{origin}).  Although we do not show the comparison, the other Gaussians also fall off with this same length.   
The $\phi$ dependence of the break-point is discussed below.
Although it is hard to capture in a graph, the full two-dimensional hopping matrix elements $|J(i,j)|$ has a block structure, with a sequence of rectangular blocks, each corresponding to a different Gaussian.



\begin{figure}[t]
\includegraphics[scale=0.6]{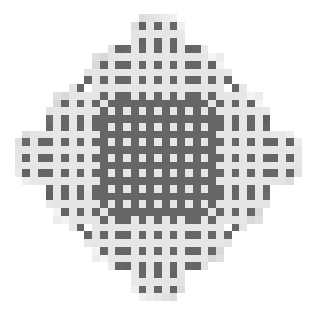}
\caption{Phases $\arg(J(0,z))$, where $J$ corresponds to hopping matrix element of
the most localized parent Hamiltonian with $\phi=4/5$. The $x$ and $y$ axis represent the locations $z=x+i y$, and the center of the figure corresponds to $z=0$. Each shaded square corresponds to a site. The dark squares represent a phase of $0$ while the light ones represent a phase of $\pi$. In the white areas phases cannot be determined due to numerical precision.  A clear block structure can be observed.}\label{fig:phases}
\end{figure}

As illustrated by Fig.~\ref{fig:phases}, the block structure also appears in the phases of the matrix elements. In that figure, we represent the phases $\arg(J(0,z))$ 
by shades of gray: lighter and darker regions correspond to phases near 0 and $\pi$.  The central region clearly agrees with the pattern in Eq.~(\ref{origin}).  The pattern is shifted in the peripheral blocks, but the periodicity is the same.
%

By systematically studying different magnetic field strengths $\phi$, we find that the block sizes grow with $\phi$.
In particular it appears that, as $\phi\to 1$, the blocks have linear dimension $s_0=1/(1-\phi)$.  Each block in the upper right quadrant can be labeled by two non-negative integers $a,b$, such that the lower left corner is at $s_0 (a + i b)$.  Within that block, the matrix elements appear to be well-approximated by
\begin{equation}
|J|\approx\exp\left(-\frac{\pi}{2}(1-\phi)(|z-\lambda|^2+s_0^2 (a+b))\right),
\end{equation}
where $\lambda=(a+i b)s_0$.  The origin of this empirical relationship is mysterious.

\section{Conclusions and Outlook}\label{conc}
Many degrees of freedom remain after requiring that a lattice model contains a flat lowest Landau level.  These degrees of freedom correspond to choosing the wavefunctions and energies of the states which are not in the flat band.  In this paper we (1) construct a subset of these Hamiltonians that have a simple structure, and (2) numerically explore the properties of the most localized Hamiltonian whith a flat lowest Landau level.

Remarkably we find that the Kapit-Mueller Hamiltonian is very close to this optimized Hamiltonian.  There are, however, small differences in some of the longer range hopping matrix elements.  In particular, the hopping elements display a remarkable block structure of elusive origin.

In all of the models we construct, the hopping matrix elements fall off as a Gaussian.  Due to this rapid decrease, an experimental implementation only needs concern itself with the largest hoppings, which are short ranged.  This speaks to the feasibility of such explorations~\cite{goldman2016topological,nielsen2013local}.  In optical lattice experiments the size of different hopping matrix elements can be tuned by adding higher harmonics to an ordinary optical lattice, or by laying out the sites in three dimensions. The NIST group has implemented the latter technology in creating a 1D lattice with tunable next-nearest neighbor hopping \cite{spielman1d}.  Implementations in superconducting circuits would require using established techniques for patterning wires which cross over one-another. The effect of the truncation will broaden the flat band, but the brodening can be optimized using the method described in \cite{PhysRevB.93.155155}.

One could imagine exploring the properties of models where we not only constrain the properties of the lowest band, but also the higher bands.  The extreme example of this is requiring that all other states are degenerate -- a case which was explored by Ataki\c{s}i and Oktel, as well as Jian, Gu and Qi \cite{oktel,jian}.  Another extension is to construct models where the flat band is spanned by the wavefunctions from the second Landau level (or higher).  The novel physics there would have to do with the different effective interactions one finds when projecting to the flat band.

\section*{Acknowledgements}
JD would like to acknowledge  Daniel Longenecker for helpful discussions.  This material is based upon work supported by the National Science Foundation under Grant No. PHY-1806357 and the ARO-MURI Non-equilibrium Many-body Dynamics Grant No. W9111NF-14-1-0003.

\end{document}